\documentclass[aps,nofootinbib,showpacs,preprint]{revtex4}

\begin{document}

\preprint{DTP-MSU/02-12} \preprint{HUTP-02/A008}
\preprint{hep-th/0204071}

\title{S-brane Solutions in Supergravity Theories}

\author{Chiang-Mei Chen}
  \email{cmchen@phys.ntu.edu.tw}
  \affiliation{Department of Physics, National Taiwan University,
Taipei 106, Taiwan, R.O.C.}

\author{Dmitri V. Gal'tsov}
 \email{galtsov@grg.phys.msu.su}
 \affiliation{Department of Theoretical Physics, Moscow State University,
119899, Moscow, Russia}

\author{Michael Gutperle}
 \email{gutperle@jacobi.physics.harvard.edu}
 \affiliation{Jefferson Physical Laboratory, Harvard University,
Cambridge, MA 02138, U.S.A.}

\date{\today}

\begin{abstract}
In this paper time dependent solutions of supergravities with
dilaton and arbitrary rank antisymmetric tensor field are found.
Although the solutions are nonsupersymmetric the equations of
motions can be integrated in a simple form. Such supergravity
solutions are related to Euclidean or spacelike branes (S-branes).
\end{abstract}

\pacs{04.20.Jb, 04.50.+h, 04.65.+e}

\maketitle

\section{Introduction}
There has been a recent surge of interest in time dependent
solutions in string theory. In \cite{KOSST01,BHKN02,Ne02,CC02}
the question of a stringy resolution of cosmological
singularities in time dependent string orbifolds was discussed.
The dS/CFT correspondence \cite{St01,BBM01} identifies time
evolution in de Sitter space with renormalization group flow
\cite{St01a}. Very recently Sen \cite{Se02a,Se02b} (see also
\cite{Gib02}) has constructed a conformal field theory
description of dynamical open string tachyon condensation. For an
earlier work on time-dependent solutions see \cite{BF94,LOW97a,
LOW97b,PS97,LW97,IM01}.

Dirichlet branes \cite{Po95} are extended solitonic objects
carrying Ramond-Ramond charge and therefore the worldvolume of
such a (static) brane includes the time direction. It is a natural
question, partly motivated by the dS/CFT correspondence, whether
there are Euclidean branes which have a purely spacelike world
volume. Euclidean branes were first constructed in
\cite{Hu98,Hu01} in type II$^*$ theories which are non-unitary
theories obtained by timelike T-duality from the standard type II
theories. The simplest starting point for the construction of a
Euclidean brane in type II theories is given by considering open
strings which satisfy Dirichlet boundary conditions in the time
direction \cite{GS02}. Such a spacelike brane (S-brane) only
exists for one instant in time.

Another argument for the existence of S-branes uses the open
string tachyons in unstable D-branes or D-brane-anti-D-brane
pairs. (Similar constructions are also possible in field theory
\cite{FGGRS95}). The basic argument for the existence of S-branes,
illustrated by a specific example, is the following. In type IIA
string theory there exists ``miss-matched'' D-branes, such as the
D3-brane, which are unstable and contain a tachyon field. Let us
consider the D3-brane as our example. The potential of the tachyon
field, $U(T)$, resembles a double well; it was argued that the
stable D2-brane is the tachyonic kink solution of the unstable D3
world volume field theory \cite{Se99}. However, one can imagine a
similar notion for the time-dependent case. Suppose the initial
data ($t=0$) for the D3-brane tachyon field is located at the
unstable maximum, $U(0)$, with a small constant positive velocity.
Then the tachyon field will roll off from the top of the potential
and evolves to the positive minimum at $t=\infty$. During this
evolution closed string radiation will be emitted and then it will
propagate to infinity. Similarly, as a consequence of time
reversal symmetry, the tachyon field will approach the negative
minimum at $t=-\infty$. This process can be realized as incoming
radiation which excites the tachyon field to the top of potential
barrier. The full picture is a timelike kink in the tachyon field
which is a S2-brane.

Using the known coupling of the spacetime RR fields to the world
volume open string tachyon it was shown that this S2-brane carries
charge, defined as the integral of the RR-field over a surrounding
sphere (including the time dimension). The same kind of charge is
carried by an ordinary D2-brane. In analogy with Sen's
identification, this timelike kink can be identified as an
SD2-brane, i.e. a Dirichlet brane arising from open string with a
Dirichlet boundary condition on the time dimension.

Obviously this construction can be generalized to other
codimensions, for example to branes as vortices in a
brane-anti-brane pair. Moreover, a similar discussion for the
initial data along the null direction will lead to null branes
(N-branes).

Both the boundary state and the tachyon picture of the S-brane
suggests that a Sp-brane (with $p+1$ dimensional Euclidean
worldvolume) in $d$ dimensions should have $ISO(p+1) \times
SO(d-p-2,1)$ symmetry. The non-compact $SO(d-p-2,1)$ can be
interpreted as the R-symmetry of a Euclidean field theory living
on the S-brane. In \cite{GS02} supergravity solutions respecting
this symmetry where found in two particular cases, it is the aim
of this paper to generalize these S-brane solutions to arbitrary
form field, codimensions and dilaton coupling. These solutions are
new interesting time dependent/cosmological solutions of
supergravities. Note however that the relation of these solutions
to the boundary state and tachyon construction of the S-brane is
quite nontrivial and not well understood at present.

\section{General S-branes}
In this section we analyze equations governing S(p-1)-branes
associated with the charge of a $q$-form field strength. The
system contains a graviton, a $q$-form field strength, F$_{[q]}$,
and a dilaton scalar, $\phi$, coupled to the form field with the
coupling constant $a$. This is a general framework which
encompasses the bosonic sector of various supergravity theories,
coming from a truncation of the low energy limit of M-theory and
string theories, by a certain choice of the dimension $d$, the
rank of the form field $q$, and the dilaton coupling $a$. In the
Einstein frame, the action is given by
\begin{equation}\label{action}
S = \int d^d x \sqrt{-g} \left( R - \frac12 \partial_\mu \phi
\partial^\mu \phi - \frac1{2\, q!} \, {\rm e}^{a\phi} \, F_{[q]}^2
\right).
\end{equation}
This action is invariant under the following discrete S-duality:
\begin{equation} \label{duality}
g_{\mu\nu} \to g_{\mu\nu}, \qquad F \to {\rm e}^{-a\phi} \ast F,
\qquad \phi \to -\phi,
\end{equation}
where $\ast$ denotes a $d$-dimensional Hodge dual. This may be
used to construct electric versions of magnetic S-branes and vice
versa. The equations of motion, derived from the variation of the
action with respect to the individual fields, are
\begin{eqnarray}
R_{\mu\nu} - \frac12 \partial_\mu \phi \partial_\nu \phi -
\frac{{\rm e}^{a\phi}}{2(q-1)!} \left[
F_{\mu\alpha_2\cdots\alpha_q} F_\nu{}^{\alpha_2\cdots\alpha_q}-
\frac{q-1}{q(d-2)} F_{[q]}^2 \, g_{\mu\nu} \right] &=& 0,
\label{Ein} \\
\partial_\mu \left( \sqrt{-g} \, {\rm e}^{a\phi} \,
F^{\mu\nu_2\cdots\nu_q} \right) &=& 0, \label{form} \\
\frac1{\sqrt{-g}}\, \partial_\mu \left( \sqrt{-g} \partial^\mu
\phi \right) - \frac{a}{2\, q!} {\rm e}^{a\phi} F_{[q]}^2 &=& 0.
\label{dil}
\end{eqnarray}

We study S-branes with a world volume given by a $p$ dimensional
conformally flat space and with a transverse space being the $k$
dimensional hyperspace $\Sigma_{k,\sigma}$ and $q-k$ dimensional
delocalized space. Obviously, in $d$ dimensions, $p=d-q-1$. With
this in mind we choose the metric
\begin{equation}\label{metric}
ds^2 = - {\rm e}^{2A} dt^2 + {\rm e}^{2B} (dx_1^2 + \cdots +
dx_p^2) + {\rm e}^{2C} \, d\Sigma_{k,\sigma}^2 + {\rm e}^{2D}
(dy_1^2 + \cdots + dy_{q-k}^2),
\end{equation}
parameterized by four $t$-dependent functions $A(t),\,B(t),\,
C(t)$ and $D(t)$. The hyperspace $\Sigma_{k,\sigma}$ for
$\sigma=0,+1,-1$ is the $k$-dimensional flat space, the sphere
and the hyperbolic space respectively. They can be described as
\begin{equation}
d\Sigma_{k,\sigma}^2 = \bar g_{ab} dz^a dz^b = \left\{
 \begin{array}{ll}
 d \psi^2 + \sinh^2\psi \, d\Omega_{k-1}^2, \qquad & \sigma=-1,\\
 d \psi^2 + \psi^2 \, d\Omega_{k-1}^2, \qquad & \sigma=0,\\
 d \psi^2 + \sin^2\psi \, d\Omega_{k-1}^2, \qquad & \sigma=+1,
 \end{array} \right.
\label{gmetric}\end{equation}
satisfying
\begin{equation}
\bar R_{ab} = \sigma (k - 1) \bar g_{ab}.
\end{equation}
The metrics above have $SO(k-1,1),\, ISO(k)$ and $SO(k)$
symmetries respectively. In \cite{GS02} in order to have a
solution with the correct R-symmetry only the case $\sigma=-1$ and
hence $SO(k-1,1)$ symmetry was considered. In the following we
will discuss all three choices of $\sigma$.

With this ansatz, the equation for the form field (\ref{form}),
can easily be solved giving
\begin{equation}\label{SolF}
F_{[q]} = b  \,\, \mbox{vol}(\Sigma_{k,\sigma}) \wedge dy_1 \wedge
\cdots \wedge dy_{q-k}.
\end{equation}
where $b$ is the field strength parameter,
$\mbox{vol}(\Sigma_{k,\sigma})$ denotes the unit volume form of
the hyperspace $\Sigma_{k,\sigma}$.

The ansatz (\ref{metric}) and (\ref{SolF}) has $q-k$ flat
directions, takes these directions to be toroidal the solutions
are in some sense `smeared' or delocalized along these directions.
Note that from tachyon picture the appearance of delocalized
coordinates is quite natural since the tachyon is localized on a
brane. The solutions of \cite{GS02} can be obtained by setting
$k=q$.

To derive the equations for the metric functions $A,\,B,\,C$ and
$D$ one calculates first the Ricci tensor for the metric
(\ref{metric}), the non-vanishing components being
\begin{eqnarray}
R_{tt} &=& - p (\ddot B + \dot B^2 - \dot A \dot B) - k (\ddot C
+ \dot C^2 - \dot A \dot C) - (q-k) (\ddot D + \dot D^2 - \dot A
\dot D), \\
R_{xx} &=& {\rm e}^{2B-2A} \left[ \ddot B - \dot A \dot B + p \dot
B^2 + k \dot B \dot C + (q-k) \dot B \dot D \right], \label{Rxx}\\
R_{yy} &=& {\rm e}^{2D-2A} \left[ \ddot D - \dot A \dot D + p \dot
B \dot D + k \dot C \dot D + (q-k) \dot D^2 \right], \label{Ryy}\\
R_{ab} &=&  \left\{ {\rm e}^{2C-2A} \left[ \ddot C - \dot A \dot C
+ p \dot B \dot C + k \dot C^2 + (q-k) \dot C \dot D \right] +
\sigma (k-1) \right\} \, \bar g_{ab}. \label{Rab}
\end{eqnarray}
\ From the expressions for the Ricci tensor, we note that the
formulation can be largely simplified once we chose the following
gauge condition
\begin{equation}\label{gauge}
- A + p B + k C + (q-k) D = 0.
\end{equation}
After taking the above gauge, the Einstein equations (\ref{Ein})
finally reduce to the following set of equations
\begin{eqnarray}
- \ddot A  + \dot A^2 - p \dot B^2 - k \dot C^2 -(q-k) \dot D^2 -
\frac12 \dot \phi^2 - \frac{(q-1)b^2}{2(d-2)} {\rm e}^{a\phi+2pB}
&=& 0, \label{EqA} \\
\ddot B + \frac{(q-1) b^2}{2(d-2)} {\rm e}^{a\phi+2pB} &=& 0,
\label{EqB} \\
\ddot C + \sigma (k-1) {\rm e}^{2A-2C} - \frac{p b^2}{2(d-2)} {\rm
e}^{a\phi+2pB} &=& 0, \label{EqC} \\
\ddot D - \frac{p b^2}{2(d-2)} {\rm e}^{a\phi+2pB} &=& 0.
\label{EqD}
\end{eqnarray}
Substituting our ansatz into the Eq.(\ref{dil}) one obtains the
following dilaton equation
\begin{equation}\label{EqPhi}
\ddot \phi + \frac{a b^2}2 \, {\rm e}^{a\phi+2pB} = 0,
\end{equation}
where dots denote derivatives with respect to $t$.

The equations (\ref{EqB}), (\ref{EqD}) and (\ref{EqPhi}) are of
similar structure, and it is easy to see that the appropriate
combinations of $D,\, B$ and $\phi,\, B$ obey simple homogeneous
equations. Therefore
\begin{equation}
\phi = \frac{a(d-2)}{q-1} B + c_1 t + c_2,
\end{equation}
with constant $c_1,\, c_2$. Similar relation can be found for $D$
and $B$, for which, however, we simply take
\begin{equation}
D = -\frac{p}{q-1} B.
\end{equation}

It is more convenient to reparameterize $A(t),\, B(t),\, C(t),\,
D(t)$ ensuring the gauge (\ref{gauge}) choice by two independent
functions $f(t),\, g(t)$ as
\begin{equation}
A = k g - \frac{p}{q-1} f, \qquad B = f, \qquad C = g -
\frac{p}{q-1} f, \qquad D = -\frac{p}{q-1} f, \label{abcdrel}
\end{equation}
consequently, the equations of motion reduce to
\begin{eqnarray}
\ddot f + \frac{(q-1)b^2}{2(d-2)} {\rm e}^{\chi f + a c_1 t + a
c_2} &=& 0, \label{Eqf} \\
\ddot g + \sigma (k-1) {\rm e}^{2(k-1) g} &=& 0, \label{Eqg} \\
\frac{p}{q-1} \ddot f - k \ddot g + k(k-1) \dot g^2 - \frac{(d-2)
\chi}{2(q-1)} \dot f^2 \qquad\qquad && \nonumber\\
- {1\over 2} c_1^2 - {a c_1 (d-2) \over (q-1)} \dot f -
\frac{(q-1) b^2}{2(d-2)} {\rm e}^{\chi f+ a c_1 t+ a c_2} &=& 0,
\label{EqMix}
\end{eqnarray}
where the parameter $\chi$ is defined as
\begin{equation}
\chi = 2 p + \frac{a^2(d-2)}{q-1}.
\end{equation}
The terms linear in $t$ can be absorbed into $f$ by defining
\begin{equation}
f(t) = h(t) - {a c_1 \over \chi} t - {a c_2\over \chi}.
\end{equation}

In terms of $h$ the equations of motion become
\begin{eqnarray}
\ddot{h} + \frac{(q-1)b^2}{2(d-2)} {\rm e}^{\chi h} &=& 0,
\label{Eqfc} \\
\ddot g + \sigma (k-1) {\rm e}^{2(k-1) g} &=& 0, \label{Eqgc} \\
\frac{p}{q-1} \ddot{h} - k \ddot g + k(k-1) \dot g^2 - \frac{(d-2)
\chi}{2(q-1)} \dot {h}^2 -{p\, c_1^2\over \chi} - \frac{(q-1)
b^2}{2(d-2)} {\rm e}^{\chi h} &=& 0. \label{EqMixc}
\end{eqnarray}
In fact, equations (\ref{Eqfc}), (\ref{Eqgc}) and (\ref{EqMixc})
are equivalent to the two first order equations
\begin{eqnarray}
\dot h^2 + \frac{(q-1) b^2}{(d-2)\chi} {\rm e}^{\chi h} &=&
\alpha^2, \\
\dot g^2 + \sigma {\rm e}^{2(k-1) g} &=& \beta^2,
\end{eqnarray}
provided the integration constants $\alpha$ and $\beta$ satisfy
\begin{equation}
{p\, c_1^2 \over \chi} + {(d-2) \chi \alpha^2\over 2(q-1)} -
k(k-1) \beta^2 = 0. \label{intrel}
\end{equation}
These equations can easily be integrated and the solution in
terms of $f$ and $g$ are given by
\begin{eqnarray}
f(t) &=&  \frac2{\chi} \ln \left( \frac{\alpha}{\cosh \left[ {\chi
\alpha \over 2}(t-t_0) \right]} \right) + \frac1{\chi} \ln \left(
{ (d-2) \chi \over (q-1) b^2 } \right) - {a c_1\over
\chi}t - {a c_2\over \chi}, \\
g(t) &=& \left\{ \begin{array}{ll}
 \frac1{k-1}\ln\left(\frac{\beta}{\sinh[(k-1) \beta (t-t_1)
   ]} \right), \qquad & \sigma=-1. \\
 \pm \beta (t-t_1), & \sigma=0. \\
 \frac1{k-1}\ln\left(\frac{\beta}{\cosh[(k-1) \beta (t-t_1)
   ]} \right), & \sigma=+1. \end{array} \right.
\end{eqnarray}
Superficially it might seem that the solution depends on six
parameters $t_0,t_1,c_1,c_2,b,\beta$. However it is possible to
eliminate two of them. Firstly $\beta$ can be eliminated by
rescaling $t\to \beta^{-1} t$ together with suitable scaling for
coordinates $\{x,y\}$, and secondly $t_1$ can be set to zero by a
shift of $t$. Hence the solution depends on four parameters.

In Table \ref{table}, we list the values of parameters for
eleven-dimensional supergravity and types IIA and IIB theories in
ten dimensions where the dilaton coupling is $a=(5-q)/2$. The
corresponding S-branes are not entirely independent, the discrete
S-duality (\ref{duality}) relates them in pairs; these
electric/magnetic pairs are indicated in parentheses.

For the S3-brane of IIB supergravity the five-form field strength
should be self-dual which is not ensured by our previous ansatz.
Therefore we solve this case separately. By self-duality
$F_{[5]}^2=0$ the equation of motion for dilaton field in the
gauge (\ref{gauge}) becomes
\begin{equation}
\ddot \phi = 0.
\end{equation}
In fact, the dilaton coupling with form field $F_{[5]}$ is absent
in IIB theory, $a=0$, so the dilaton field can be set to a
constant. Following an analogous calculation we found that the
self-dual five-form field should be
\begin{equation}
F_{[5]} = \frac{b}{\sqrt2} (1 + \ast)  \,\,
\mbox{vol}(\Sigma_{k,\sigma}) \wedge dy_1 \wedge \cdots \wedge
dy_{5-k}.
\end{equation}
The S3-brane solution is therefore given by setting $a=0$ and
$c_1=0$ and the metric can be directly read from the general
expressions of solutions given in this section by using the values
of parameters in Table \ref{table}.

\begin{table}
\caption{\label{table} Parameters of S-branes}
\begin{tabular}{|c||c|c||c|c|c|c|c|c|c|c|c|} \hline
  & \multicolumn{2}{|c|}{M-theory}
  & \multicolumn{9}{|c|}{Type II string theories} \\
  \hline
  & S5 & S2 & S6 & NS S5 & S5 & S4 & [S3] & S2 ($\ast$S4) &
    S1 ($\ast$S5)& NS S1 ($\ast$NS S5) & S0 ($\ast$S6) \\
  \hline\hline
  d & 11 & 11 & 10 & 10 & 10 & 10 & 10 & 10 & 10 & 10 &10 \\
  \hline
  $q$ & 4 & 7 & 2 & 3 & 3 & 4 & 5 & 6 & 7 & 7 & 8 \\
  \hline
  $a$ & 0 & 0 & 3/2 & -1 & 1 & 1/2 & 0 & -1/2 &
  -1 & 1 &-3/2 \\
  \hline\hline
  $p$ & 6 & 3 & 7 & 6 & 6 & 5 & 4 & 3 & 2 & 2 &1 \\
  \hline
  $\chi$ & 12 & 6 & 32 & 16 & 16 & 32/3 & 8 &
  32/5 & 16/3 & 16/3 & 32/7 \\
  \hline
\end{tabular}
\end{table}

\subsection{Hyperbolic transverse space}
In this subsection we will discuss the form of the metric in
special limiting cases. In order to simplify notation we set
$t_1=0$ and $\beta=1$ by a shift and rescaling discussed before.

The asymptotic region is at $t\to 0$ where the radius of the
$\Sigma_{k,-1}$ diverges. Defining $u=[(k-1)t]^{-1/(k-1)}$, near
$t=0, u=\infty$ the metric becomes
\begin{equation}
ds^2_{t\to 0} \sim {\rm e}^{-\frac{2p f_0}{q-1}} ( - du^2 + u^2
d\Sigma_{k,-1}^2 + d\vec y_{q-k}^2 ) + {\rm e}^{2f_0} d\vec x_p^2,
\end{equation}
with
\begin{equation}
f_0 = \frac2{\chi} \ln \left( \frac{\alpha}{\cosh \left( {\chi
\alpha \over 2} t_0 \right)} \right) + \frac1{\chi} \ln \left( {
(d-2) \chi \over (q-1) b^2 } \right) - {a c_2\over \chi}.
\end{equation}

The large $t$, near-brane behavior is given by
\begin{equation}
ds^2_{t\to \infty} \sim {\rm e}^{-\frac{2pf_1}{q-1}} {\rm
e}^{\frac{2p}{q-1}(\alpha+\frac{ac_1}{\chi})t} \left( -
2^{\frac{2k}{k-1}} {\rm e}^{-kt} dt^2 + 2^{\frac2{k-1}} {\rm
e}^{-t} d\Sigma_{k,-1}^2 + d\vec y_{q-k}^2 \right) + {\rm
e}^{2f_1} {\rm e}^{-2(\alpha+\frac{ac_1}{\chi})t} d\vec x_p^2,
\end{equation}
with
\begin{equation}
f_1 = \frac2{\chi} \ln \alpha + \alpha t_0 - \frac2{\chi} \ln 2 +
\frac1{\chi} \ln \left( { (d-2) \chi \over (q-1) b^2 } \right) -
{a c_2\over \chi}.
\end{equation}
Even thought the Ricci scalar tends to zero in this region the
geometry is singular because for example the coefficient of
$d\vec x_{p}^2$ vanishes.

\subsection{Flat transverse space}
The asymptotic region near $t \to 0$ the metric becomes
\begin{equation}
ds^2_{t\to 0} \sim {\rm e}^{-\frac{2p f_0}{q-1}} ( - dt^2 +
d\Sigma_{k,0}^2 + d\vec y_{q-k}^2 ) + {\rm e}^{2f_0} d\vec x_p^2.
\end{equation}

The large $t$, near-brane behavior is given by
\begin{equation}
ds^2_{t\to \infty} \sim {\rm e}^{-\frac{2pf_1}{q-1}} {\rm
e}^{\frac{2p}{q-1}(\alpha+\frac{ac_1}{\chi})t} \left( - {\rm
e}^{\pm 2kt} dt^2 + {\rm e}^{\pm 2t} d\Sigma_{k,0}^2 + d\vec
y_{q-k}^2 \right) + {\rm e}^{2f_1} {\rm
e}^{-2(\alpha+\frac{ac_1}{\chi})t} d\vec x_p^2.
\end{equation}

\subsection{Spherical transverse space}
The asymptotic region near $t \to 0$ the metric becomes
\begin{equation}
ds^2_{t\to 0} \sim {\rm e}^{-\frac{2p f_0}{q-1}} ( - dt^2 +
d\Sigma_{k,+1}^2 + d\vec y_{q-k}^2 ) + {\rm e}^{2f_0} d\vec x_p^2.
\end{equation}

The large $t$, near-brane behavior is given by
\begin{equation}
ds^2_{t\to \infty} \sim {\rm e}^{-\frac{2pf_1}{q-1}} {\rm
e}^{\frac{2p}{q-1}(\alpha+\frac{ac_1}{\chi})t} \left( -
2^{\frac{2k}{k-1}} {\rm e}^{-kt} dt^2 + 2^{\frac2{k-1}} {\rm
e}^{-t} d\Sigma_{k,+1}^2 + d\vec y_{q-k}^2 \right) + {\rm
e}^{2f_1} {\rm e}^{-2(\alpha+\frac{ac_1}{\chi})t} d\vec x_p^2.
\end{equation}

\section{Static Solutions}
In this section we will  briefly describe the application of the
ansatz and gauge we used in the previous section to the case of
static solutions as it turns out these solutions are related (for
a different choice of gauge) to the general black brane
\cite{HoSt91} solutions found in \cite{ZZ99}, (see also
\cite{BMO01,GQZT01}).

The ansatz for static solution is given by
\begin{equation}
ds^2 = {\rm e}^{2A} dr^2 + {\rm e}^{2B} (- dt^2 + dx_1^2 + \cdots
+ dx_{p-1}^2) + {\rm e}^{2C} \, d\Sigma_{k,\sigma}^2 + {\rm
e}^{2D} (dy_1^2 + \cdots + dy_{q-k}^2).
\end{equation}
With the same metric for $\Sigma_{k,\sigma}$ given by
(\ref{gmetric}) and field strength (\ref{SolF}) but in this case
all functions $A(r),\,B(r),\,C(r)$ and $D(r)$ depend only on the
radius coordinate $r$. Using the gauge condition (\ref{gauge})
the equations of motion become (where primes now denote
derivatives with respect to $r$).
\begin{eqnarray}
- A''  + A'^2 - p B'^2 - k C'^2 - (q-k) D'^2 - \frac12 \phi'^2 +
\frac{(q-1)b^2}{2(d-2)} {\rm e}^{a\phi+2pB} &=& 0, \label{EqAd}\\
B'' - \frac{(q-1) b^2}{2(d-2)} {\rm e}^{a\phi+2pB} &=& 0,
\label{EqBd} \\
C'' - \sigma (k-1) {\rm e}^{2A-2C} + \frac{p b^2}{2(d-2)} {\rm
e}^{a\phi+2pB} &=& 0, \label{EqCd} \\
D'' + \frac{p b^2}{2(d-2)} {\rm e}^{a\phi+2pB} &=& 0. \label{EqDd}
\end{eqnarray}
Again $\phi$ must be related to the function $B$ as follows
\begin{equation}\label{SolPhib}
\phi = \frac{a(d-2)}{q-1} B + c_1 r + c_2.
\end{equation}
Using the same relations as in (\ref{abcdrel}) the equations of
motion can be reduced to two first order differential equations
for $f(r),\,g(r)$. The solutions are given by
\begin{eqnarray}
f(r) &=& \frac2{\chi} \ln \left( \frac{\alpha}{\sinh \left[ {\chi
\alpha \over 2}(r-r_0) \right]} \right) + \frac1{\chi} \ln \left(
{ (d-2) \chi \over (q-1) b^2 } \right) - {a c_1 \over
\chi}r - {a c_2\over \chi} , \\
g(r) &=& \left\{ \begin{array}{ll}
 \frac1{k-1}\ln\left(\frac{\beta}{\cosh[(k-1) \beta (r-r_1)
   ]} \right), \qquad & \sigma=-1. \\
 \pm \beta (r-r_1), & \sigma=0. \\
 \frac1{k-1}\ln\left(\frac{\beta}{\sinh[(k-1) \beta (r-r_1)
   ]} \right), & \sigma=+1. \end{array} \right.
\end{eqnarray}
After rescaling and shifts the solution will depend on four
parameters ($r_0,c_1,c_2,b$). It is instructive to compare the
$\sigma=+1$ case to the fully localized, $k=q$, three-parameter
($\rho_0,\bar c_1,\bar c_2$) solutions found in \cite{ZZ99,BMO01}
for type II theories in ten dimensions.\footnote{
The most general solutions in \cite{BMO01} actually contain a
fourth parameter $\bar c_3$ which seems not relate to the
parameter $c_2$ here. Moreover, please also note our notation of
$p$ has value one different from the convention in \cite{BMO01}.}
A closer analysis shows that these solutions are indeed
equivalent after a coordinate transformation
\begin{equation}
r = \frac1{8-p} \ln \left(
\frac{1+(\rho_0/\rho)^{8-p}}{1-(\rho_0/\rho)^{8-p}} \right),
\end{equation}
and specific value of $c_2$
\begin{equation}
c_2 = \frac{4(p-4)}{p(8-p)} \ln \left( \frac{b}{\kappa(8-p)}
\sinh[(p-8)\kappa r_0] \right).
\end{equation}
The relation of parameters is
\begin{eqnarray}
\bar c_1 &=& - \frac{c_1}{8-p}, \\
\bar c_2 &=& \coth[(p-8) \kappa r_0], \\
\rho_0 &=& 2^{1/(p-8)} \exp \left[ \frac{p}{4(p-4)} c_2 \right],
\end{eqnarray}
where
\begin{equation}
\kappa^2 = \frac{2(9-p)}{8-p} - \frac{p \, c_1^2}{16(8-p)}.
\end{equation}

In \cite{BMO01} the static solutions for $\sigma=+1$ where
interpreted as supergravity solutions corresponding to coincident
brane/anti-brane pairs. Note that whether this interpretation is
correct is not clear a priori since one would not expect to have
a static time independent solution for an object which is
unstable and decays. It is however tempting to speculate that the
time dependent solutions we have found could describe exactly
such an process.

\section{Conclusion}
In this paper we have constructed new time dependent solutions in
supergravities. For transverse spaces which are hyperbolic these
solutions generalize the ones found in \cite{GS02} to arbitrary
codimension, rank of field strength and dilaton coupling. These
solutions are expected to be supergravity realizations of
S-branes, Euclidean branes which only exist at an instant in
time. Although the solutions are not supersymmetric the field
equations can be integrated (for the hyperbolic as well as the
flat and spherical case). One motivation for considering S-branes
was the role which Euclidean branes play in the dS/CFT
correspondence and the role of holography in comparison to AdS/CFT
\cite{Ma98,GKP98,Wi98,AGMOO99}. It would be very interesting to
explore the role the solutions in this paper might play in this
context. Relatedly it is an interesting question whether the
solutions in this paper have a cosmological interpretation and if
an S-brane can be used to get a nonsingular connection between big
crunch and big bang cosmologies.

The same gauge and ansatz can be used to find static solutions. We
showed that these solutions are equivalent to the ones found in
\cite{ZZ99,BMO01}. We have speculated that the time dependent
solutions could be realizations of an brane anti-brane
annihilation process it would be very interesting to explore this
relation further. Furthermore given the relation of brane
antibrane systems to fluxbranes \cite{Me64,GM88,GW87,DGGH95,Ts95,
DGGH96,RuTs96,CG00,CGS99,Sa01,GS01,CHC01,CGS02,Iv02}, it might be
possible that the time dependent solutions describe the dynamical
evolution of fluxbranes. We leave this question for future work.

\begin{acknowledgments}
We would like to thank Gerard Clement, Pei-Ming Ho, Andy
Strominger, John Wang and Hyun Seok Yang for discussions. The
work of CMC was supported by the CosPA project, the National
Science Council, Taiwan, and in part by the Center of Theoretical
Physics at NTU and the National Center for Theoretical Sciences.
The work of DVG was supported by the Russian Foundation for Basic
Research under grant 00-02-16306. The work of MG was supported by
DOE grant DE-FG02-91ER40655.
\end{acknowledgments}



\begin{thebibliography}{99}

\bibitem{KOSST01}
    J. Khoury, B. A. Ovrut, N. Seiberg, P. J. Steinhardt and N. Turok,
    {\sl From big crunch to big bang},
    {\tt hep-th/0108187}.

\bibitem{BHKN02}
    V. Balasubramanian, D. F. Hassan, E. Keski-Vakkuri and A. Naqvi,
    {\sl A space-time orbifold: a toy model for a cosmological singularity},
    {\tt hep-th/0202187}.

\bibitem{Ne02}
    N. A. Nekrasov,
    {\sl Milne universe, tachyons, and quantum group},
    {\tt hep-th/0203112}.

\bibitem{CC02}
    L. Cornalba and M. S. Costa,
    {\sl A new cosmological scenario in string theory},
    {\tt hep-th/0203031}.

\bibitem{St01}
    A. Strominger,
    {\sl The dS/CFT correspondence},
    {\it JHEP \bf 0110} (2001) 034;
    {\tt hep-th/0106113}.

\bibitem{BBM01}
    V. Balasubramanian, J. de Boer and D. Minic,
    {\sl Mass, entropy and holography in asymptotically de Sitter spaces},
    {\tt hep-th/0110108}.

\bibitem{St01a}
    A. Strominger,
    {\sl Inflation and the dS/CFT correspondence},
    {\it JHEP \bf 0111} (2001) 049;
    {\tt hep-th/0110087}.

\bibitem{Se02a}
    A. Sen,
    {\sl Rolling tachyon},
    {\tt hep-th/0203211}.

\bibitem{Se02b}
    A. Sen,
    {\sl Tachyon matter},
    {\tt hep-th/0203265}.

\bibitem{Gib02}
    G. W. Gibbons,
    {\sl Cosmological evolution of the rolling tachyon},
    {\tt hep-th/0204008}.

\bibitem{BF94}
    K. Behrndt and S. F\"{o}rste,
    {\sl String-Kaluza-Klein cosmology},
    {\it Nucl. Phys. \bf B430} (1994) 441-495;
    {\tt hep-th/9403179}.

\bibitem{LOW97a}
    A. Lukas, B. A. Ovrut and D. Waldram,
    {\sl Cosmological solutions of type II string theory},
    {\it Phys. Lett. \bf B393} (1997) 65-71;
    {\tt hep-th/9608195}.

\bibitem{LOW97b}
    A. Lukas, B. A. Ovrut and D. Waldram,
    {\sl String and M-theory cosmological solutions with Ramond forms},
    {\it Nucl. Phys. \bf B495} (1997) 365-399;
    {\tt hep-th/9610238}.

\bibitem{PS97}
    R. Poppe and S. Schwager,
    {\sl String Kaluza-Klein cosmologies with RR-fields},
    {\it Phys. Lett. \bf B393} (1997) 51-58;
    {\tt hep-th/9610166}.

\bibitem{LW97}
    F. Larsen and F. Wilczek,
    {\sl Resolution of cosmological singularities},
    {\it Phys. Rev. \bf D55} (1997) 4591-4595;
    {\tt hep-th/9610252}.

\bibitem{IM01}
    V. D. Ivashchuk and V. N. Melnikov,
    {\sl Exact solutions in multidimensional gravity with antisymmetric
         forms, topical review},
    {\it Class. Quantum Grav. \bf 18} (2001) R87-R152;
    {\tt hep-th/0110274}.

\bibitem{Po95}
    J. Polchinski,
    {\sl Dirichlet-branes and Ramond-Ramond charges},
    {\it Phys. Rev. Lett. \bf 75} (1995) 4724-4727;
    {\tt hep-th/9510017}.

\bibitem{Hu98}
    C. M. Hull,
    {\sl Timelike T-duality, de Sitter space, large N gauge theories
         and topological field theory},
    {\it JHEP \bf 9807} (1998) 021;
    {\tt hep-th/9806146}.

\bibitem{Hu01}
    C. M. Hull,
    {\sl de Sitter space in supergravity and M theory},
    {\it JHEP \bf 0111} (2001) 012;
    {\tt hep-th/0109213}.

\bibitem{GS02}
    M. Gutperle and A. Strominger,
    {\sl Spacelike branes},
    {\tt hep-th/0202221}.

\bibitem{FGGRS95}
    E. Farhi, J. Goldstone, S. Gutmann, K. Rajagopal and R. J. Singleton,
    {\sl Fermion production in the background of Minkowski space classical
         solutions in spontaneously broken gauge theory},
    {\it Phys. Rev. \bf D51} (1995) 4561;
    {\tt hep-ph/9410365}.

\bibitem{Se99}
    A. Sen,
    {\sl Non-BPS states and branes in string theory},
    {\tt hep-th/9904207}.

\bibitem{ZZ99}
    B. Zhou and C.-J. Zhu,
    {\sl The complete black brane solutions in d-dimensional
         coupled gravity system},
    {\tt hep-th/9905146}.

\bibitem{BMO01}
    P. Brax, G. Mandal and Y. Oz,
    {\sl Supergravity description of Non-BPS branes},
    {\it Phys. Rev. \bf D63} (2001) 064008;
    {\tt hep-th/0005242}.

\bibitem{HoSt91}
    G. T. Horowitz and A. Strominger,
    {\sl Black strings and p-branes},
    {\it Nucl. Phys. \bf B360} (1991) 197-209.

\bibitem{GQZT01}
    C. Grojean, F. Quevedo, I. Zavala and G. Tasinato,
    {\sl Branes on charged dilatonic background: self-tuning, Lorentz
         violations and cosmology},
    {\it JHEP \bf 0108} (2001) 005;
    {\tt hep-th/0106120}.

\bibitem{Ma98}
    J. Maldacena,
    {\sl The large $N$ limit of superconformal field theories and
         supergravity},
    {\it Adv. Theor. Math. Phys. \bf 2} (1998) 231-252,
    {\it Int. J. Theor. Phys. \bf 38} (1999) 1113-1133;
    {\tt hep-th/9711200}.

\bibitem{GKP98}
    S. S. Gubser, I. R. Klebanov and A. M. Polyakov,
    {\sl Gauge theory correlators from non-critical string theory},
    {\it Phys. Lett. \bf B428} (1998) 105-114;
    {\tt hep-th/9802109}.

\bibitem{Wi98}
    E. Witten,
    {\sl Anti-de Sitter space and holography},
    {\it Adv. Theor. Math. Phys. \bf 2} (1998) 253-291;
    {\tt hep-th/9802150}.

\bibitem{AGMOO99}
    A. Aharony, S. S. Gubser, J. Maldacena, H. Ooguri and Y. Oz,
    {\sl Large $N$ field theories, string theory and gravity},
    {\it Phys. Rept. \bf 323} (2000) 183-386;
    {\tt hep-th/9905111}.

\bibitem{Me64}
    M. A. Melvin,
    {\sl Pure magnetic and electric geons},
    {\it Phys. Lett. \bf 8} (1964) 65.

\bibitem{GM88}
    G. W. Gibbons and K. Maeda,
    {\sl Black holes and membranes in higher dimensional theories
         with dilaton fields},
    {\it Nucl. Phys. \bf B298} (1988) 741-775.

\bibitem{GW87}
    G. W. Gibbons and D. L. Wiltshire,
    {\sl Space-time as a membrane in higher dimensions},
    {\it Nucl. Phys. \bf B287} (1987) 717;
    {\tt hep-th/0109093}.

\bibitem{DGGH95}
    H. F. Dowker, J. P. Gauntlett, G. W. Gibbons and G. T. Horowitz,
    {\sl Decay of magnetic fields in Kaluza-Klein theory},
    {\it Phys. Rev. \bf D52} (1995) 6929-6940;
    {\tt hep-th/9507143}.

\bibitem{DGGH96}
    H. F. Dowker, J. P. Gauntlett, G. W. Gibbons and G. T. Horowitz,
    {\sl Nucleation of p-branes and fundamental strings},
    {\it Phys. Rev. \bf D53} (1996) 7115-7128;
    {\tt hep-th/9512154}.

\bibitem{Ts95}
    A. A. Tseytlin,
    {\sl Melvin solution in string theory},
    {\it Phys. Lett. \bf B346} (1995) 55;
    {\tt hep-th/9411198}.

\bibitem{RuTs96}
    J. G. Russo and A. A. Tseytlin,
    {\sl Magnetic flux tube models in superstring theory},
    {\it Nucl. Phys. \bf B461} (1996) 131;
    {\tt hep-th/9508068}.

\bibitem{CG00}
    M. S. Costa and M. Gutperle,
    {\sl The Kaluza-Klein Melvin solution in M-theory},
    {\it JHEP \bf 0103} (2001) 027;
    {\tt hep-th/0012072}.

\bibitem{CGS99}
    C.-M. Chen, D. V. Gal'tsov and S. A. Sharakin,
    {\sl Intersecting M-fluxbranes},
    {\it Grav. Cosmol. \bf 5} (1999) 45-48;
    {\tt hep-th/9908132}.

\bibitem{Sa01}
    P.M. Saffin,
    {\sl Gravitating fluxbranes},
    {\it Phys. Rev. \bf D64} (2001) 024014;
    {\tt gr-qc/0104014}.

\bibitem{GS01}
    M. Gutperle and A. Strominger,
    {\sl Fluxbranes in string theory},
    {\it JHEP \bf 0106} (2001) 035;
    {\tt hep-th/0104136}.

\bibitem{CHC01}
    M. S. Costa, C. A. R. Herdeiro and L. Cornalba,
    {\sl Flux-branes and the dielectric effect in string theory},
    {\it Nucl. Phys. \bf B619} (2001) 155-190;
    {\tt hep-th/0105023}.

\bibitem{CGS02}
    C.-M. Chen, D. V. Gal'tsov and P. M. Saffin,
    {\sl Supergravity fluxbranes in various dimensions},
    {\it Phys. Rev. \bf D65} (2002) 084004;
    {\tt hep-th/0110164}.

\bibitem{Iv02}
    V. D. Ivashchuk,
    {\sl Composite fluxbranes with general intersections},
    {\tt hep-th/0202022}.

\end{thebibliography}
\end{document}